\documentclass[prl, twocolumn,amsmath]{revtex4}
\usepackage{graphicx}
\usepackage{dcolumn}
\usepackage{bm}
\usepackage{mathbbol}

\newcommand{\mathcommand}[3][0]{\newcommand{#2}[#1]{\ensuremath{#3}}}

\newcommand{\ts}[2]{{#1}_{\textnormal{#2}}} 
\newcommand{\tsc}[2]{{#1}_{\textsc{#2}}} 

\newcommand{\abbrev}[1]{{\scshape\lowercase{#1}}}  

\newcommand{\be}{\begin{equation}}
\newcommand{\ee}{\end{equation}}
\newcommand{\splitstart}{\begin{equation}\begin{split}}
\newcommand{\splitstop}{\end{split}\end{equation}} 

\newcommand{\refeq}[1]{Eq.~\eqref{#1}}

\newcommand{\reffig}[1]{Fig.~\ref{#1}}

\renewcommand{\Im}[1]{\ensuremath{
    \text{Im}\left\{#1\right\}}}
\renewcommand{\Re}[1]{\ensuremath{
    \text{Re}\left\{#1\right\}}}

\newcommand{\ham}[1][]{\ensuremath{{H}_{\text{#1}}}} 
\newcommand{\tildeham}[1][]{\ensuremath{{\tilde{H}}_{\text{#1}}}}

\mathcommand[1]{\smallket}{|#1\rangle}
\mathcommand[1]{\smallbra}{\langle #1|}
\mathcommand[1]{\bigket}{\bigl|#1\bigr\rangle}
\mathcommand[1]{\bigbra}{\bigl\langle #1\bigr|}
\mathcommand[1]{\biggket}{\biggl|#1\biggr\rangle}
\mathcommand[1]{\biggbra}{\biggl\langle #1\biggr|}
\mathcommand[1]{\ket}{\left| #1 \right\rangle}  
\mathcommand[1]{\lket}{\bigl| #1 \bigr)}  
\mathcommand[1]{\bra}{\left\langle #1 \right|}  
\mathcommand[1]{\lbra}{\bigl( #1 \bigr|}  
\newcommand{\ketbra}[2]{\left| #1 \right\rangle\!\!\left\langle #2
\right|} 

\newcommand{\nodag}{{\phantom{\dag}}} 
\newcommand{\nostar}{{\phantom{*}}} 

\newcommand{\Tr}[2][]{\text{Tr}_\text{#1}\left\{#2\right\}}

\newcommand{\latinfont}[1]{#1}  

\newcommand{\etal}{\latinfont{et al.}} 

\newcommand\qu[1]{``#1"}
\newcommand{\acs}{\abbrev{ac}-Stark shift}
\mathcommand{\te}{\text{e}}
\mathcommand{\vactext}{\text{vac}}
\mathcommand{\vacket}{\ket\vactext}

\newcommand{\QD}{\abbrev{QD}}
\newcommand{\QDs}{\abbrev{QD}s}

\begin{document}
\title{A simple, robust, and scalable quantum logic gate with quantum dot cavity QED systems}
\author{T. D. Ladd}
\email[Electronic address: ]{tdladd@gmail.com}
\author{Y. Yamamoto}
\affiliation{
    Edward L. Ginzton Laboratory,
    Stanford University,
    Stanford, California 94305-4088, USA
} \affiliation{
    National Institute of Informatics,
    2-1-2 Hitotsubashi, Chiyoda-ku,
    Tokyo 101-8430, Japan
}

\begin{abstract}
We present a method to enact a deterministic, measurement-free, optically generated controlled-phase gate on two qubits defined by single electrons trapped in large-area quantum dots in a planar microcavity.  This method is robust to optical quantum dot inhomogeneity, requires only a modest-$Q$ planar cavity, employs only a single laser pulse, and allows the integration of many entangled qubits on one semiconductor chip.  We present the gate in the contexts of both adiabatic evolution and geometric phases, and calculate the degradation of performance in the presence of both spontaneous emission and cavity loss.
\end{abstract}
\maketitle

Quantum computers offer potential speed-ups over classical computers for certain problems, but only if they are made sufficiently large, fast, and robust.  Optically controlled quantum dots (\QDs) are promising for this application due to their capabilities for ultrafast optical initialization, control, and measurement \cite{shamU1,flattespinrot,fast_rotations_prop,awschsingle,pressnature} as well as their strong interface to single photons \cite{DotStrongCoupling,fushman08}.  In the last decade, numerous proposals for entangling \QDs\ with photons enhanced by optical cavities have emerged; for a few examples, see Refs.~\onlinecite{IABDLSS99,dk04,geometric_dots_zanardi,yls05,polariton_mediate}.  However, most of these schemes fail when experimentally realistic values of the cavity quality factor $Q$ and the vacuum Rabi splitting $g$ are considered.   Even for experimental parameters corresponding to the most heroic cavity quantum electrodynamics (\abbrev{QED}) demonstrations, scaling these schemes may introduce unrealistic requirements for \QD\ tuning and placement, or extremely challenging optical control sequences.

Here, we argue that an entangling logic gate between neighboring \QD\ spins is possible in a planar microcavity using only a single pulse of laser light.  Moreover, the gate we propose functions when the \QDs\ have very different optical resonances, and without requiring
the strong coupling regime of cavity \abbrev{QED}.  In conjunction with single-qubit initialization, control, and detection techniques, as well as the introduction of high-threshold, nearest-neighbor-only techniques for topological fault-tolerance in cluster states~\cite{rh07}, this gate offers strong promise for a fast, scalable fault-tolerant quantum information processing system using \QDs~\cite{ijqi}.

The single-pulse entangling gate may be understood as a
consequence of the adiabatic theorem: a quantum system in an
eigenstate remains in that eigenstate if its Hamiltonian does
not change \qu{too quickly.}  For optical control, \qu{too
quickly} means that the duration of the optical pulse should be
longer than the inverse of the optical detuning, which may
easily be large enough to keep the adiabatic condition even
with picosecond pulses.  In large-area \QDs, this ultrafast
control is enabled by the large, mesoscopically enhanced
oscillator strength of an
exciton~\cite{bpjy94,bonadeo00,guest02}, which is enhanced by a
factor of $(a/\tsc{a}{b}^*)^2$ in comparison to that of an
atom, where $a$ is the \QD\ radius and $\tsc{a}{b}^*$ is the
Bohr radius~\cite{bpjy94}.

\begin{figure}
\begin{center}
\includegraphics[width=0.8\columnwidth]{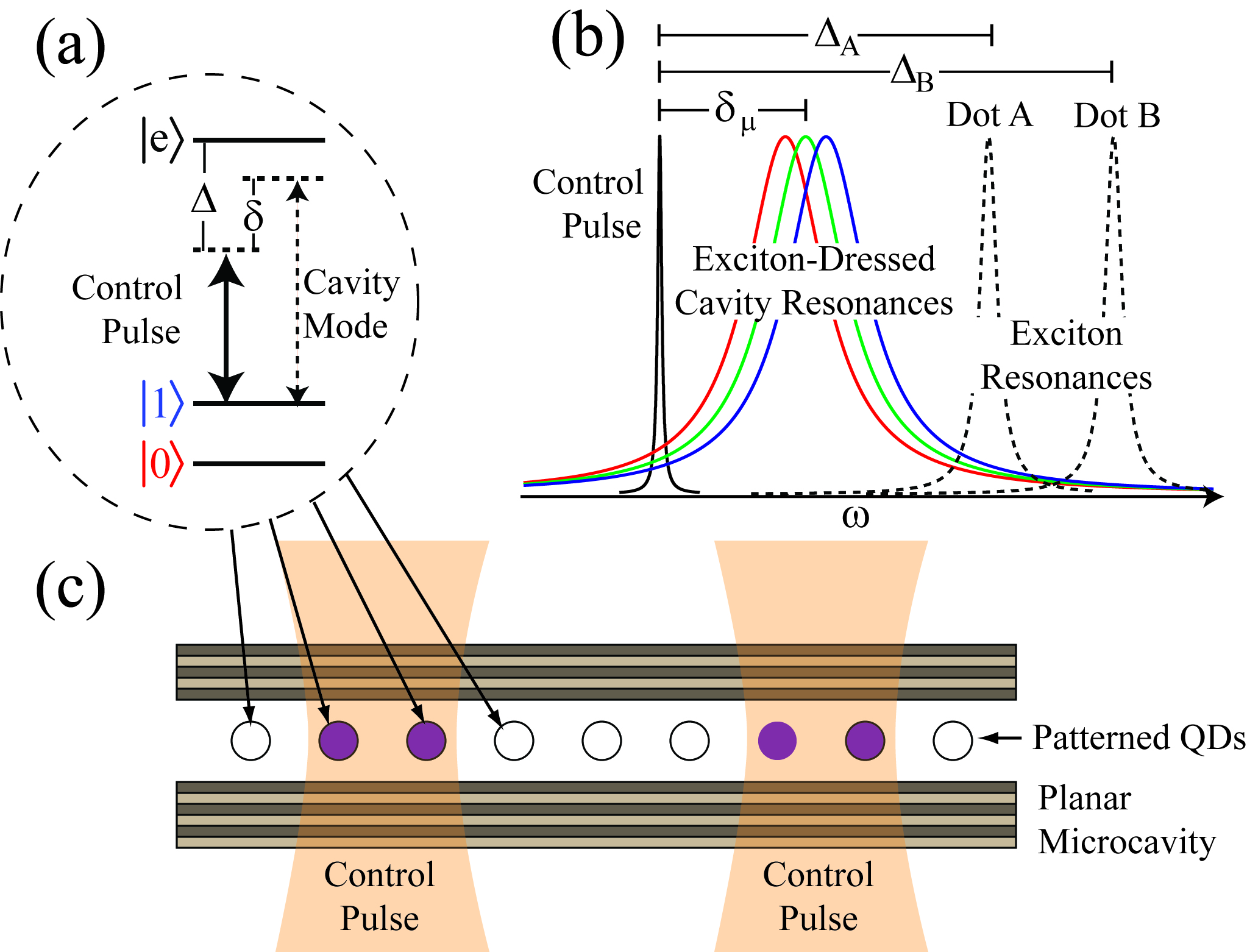}
\end{center}
\caption{(Color online) System schematic. (a) The energy levels of the quantum dot; level spin-ground-state $\ket{1}$ is coupled to trion state $\ket{e}$ by a cavity mode and control pulse.  (b) The cavity, pulse, and dot spectra; the pulse is red-detuned from the cavity, and the dot's exciton resonances are blue-detuned.  The cavity response is itself shifted by the presence of excitonic transitions, which depend on the qubit states.  (c) Optical pulses are focused to overlap two neighboring patterned quantum dots in the planar microcavity.}
\label{levels}
\end{figure}

As shown in \reffig{levels}(a), a single spin in a \QD\ has two
Zeeman-split ground states (either an electron spin-1/2 or the
$J=\pm 3/2$ states of a trapped hole.) However, circularly
polarized light connects only one of these levels, which we
notate $\tsc{\ket{1}}{a}$, to an optically excited trion state,
which we notate $\tsc{\ket\te}{a}$.  The other ground state,
labeled $\tsc{\ket{0}}{a}$ and with energy
$-\hbar\tsc\omega{a}$, is \qu{dark}, and so is unaffected by
the pulse. Neglecting the possibility of a cavity for the time
being (and so with $\delta=0$), when an optical pulse of
amplitude $\tsc\Omega{a}(t)$ is introduced with detuning
$\tsc\Delta{a}$, the semiclassical Hamiltonian in the rotating
reference frame of the optical pulse for qubit \textsc{a}
changes as
\be
\label{qubitham}
\ham[\textsc{a}](t)=-\tsc\omega{a}\tsc{\ketbra{0}{0}}{a}
                    +\tsc\Delta{a}\tsc{\ketbra{\te}{\te}}{a}
                    +\tsc\Omega{a}^\nostar(t)\tsc{\ketbra{\te}{1}}{a}
                    +\tsc\Omega{a}^*(t)\tsc{\ketbra{1}{\te}}{a}
\ee
The adiabatic theorem tells us that a system beginning in state
$\tsc{\ket{1}}{a}$ remains in state $\tsc{\ket{1}}{a}$; however
this eigenstate is dynamically \qu{dressed} by the
field pulse into a state with reduced energy in comparison to
the bare ground state.  The resulting phase shift of this state
---the \acs--- with respect to the dark state accomplishes a
single qubit
rotation~\cite{shamU1,flattespinrot,fast_rotations_prop,awschsingle,pressnature}
by an angle $\tsc\theta{a}$. The total angle $\tsc\theta{a}$
accomplished over the duration of the pulse is $
\tsc{\theta}{a} = (\tsc{\Delta}{a}/{2}) \int_{-\infty}^\infty
\bigl[\sqrt{1+{4\tsc{\Omega}{a}^2(t)}/{\tsc{\Delta}{a}^2}}-1\bigr]dt.
$ The data in Fig.~1b of Ref.~\onlinecite{pressnature} is well
described by this equation, indicating adiabatic evolution.

To move to a two-qubit-gate, we now add a second qubit and an optical
cavity.
The optical laser pulse couples to cavity modes indexed by $\mu$; each mode $\mu$ is detuned from the laser by $\delta_\mu$ and annihilated by $b_\mu$.   The part of the pulse coupling to mode $\mu$ has
(semi-classical) envelope $F_\mu(t)$ and coupling rate $\chi_\mu$, so the Hamiltonian for
the optical pulse and the empty cavity is
\be
\ham[optical](t)= \sum_{\mu} \delta_\mu a_\mu^\dag a_\mu^\nodag
+\chi_\mu F^\nostar_\mu(t) a_\mu^\dag + \chi_\mu^*
F_\mu^*(t)a_\mu^\nodag.
\ee
Adding optical loss from cavity mode $\mu$ at rate
$\kappa_\mu$, an empty cavity would evolve by this Hamiltonian
with state $\prod_\mu D_\mu[\alpha_\mu(t)]\vacket,$ for
displacement operator $D_\mu[\alpha_\mu(t)]=\exp[\alpha_\mu(t)
a_\mu^\dag - \alpha_\mu(t) a_\mu]$ and
$\alpha_\mu(t)=-i\int_0^t F_\mu(\tau)
\exp[(\tau-t)(i\delta_\mu+\kappa_\mu/2)] d\tau.$
As a first step of our quantum treatment, we work in an
interaction picture where these dynamics
are removed; i.e. we move from a density matrix $\rho(t)$ in
the rotating reference frame to $\tilde\rho(t)$ in a new
reference frame with the definition $\rho(t)=\prod_\mu
D_\mu^\nodag[\alpha_\mu(t)] \tilde\rho(t) \prod_\nu
D_\nu^\dag[\alpha_\nu(t)].$  This is a time-dependent basis
change, and so it introduces new time-dependent terms in our
Hamiltonian; in particular the cavity \abbrev{qed} Hamiltonian
becomes
\be
\begin{split}
\ham[\textsc{cqed}]&=\sum_{j=\textsc{a,b}}\sum_\mu g_{j\mu}
\ketbra{\te}{1}_j a_\mu + \text{h.c.} \rightarrow\\
\tildeham[\textsc{cqed}](t) &=
    \sum_{j=\textsc{a,b}}\sum_\mu g_{j\mu}
    \ketbra{\te}{1}_j [a_\mu+\alpha_\mu(t)] + \text{h.c.},
\end{split}
\ee
where h.c. is hermitian conjugate. If we now define
$\Omega_j(t)=\sum_\mu g_{j\mu}\alpha_\mu(t)$, then in this
reference frame we have a cavity \abbrev{qed} system with an
additional semiclassical driving term as already introduced in
\refeq{qubitham}, so the full Hamiltonian is $\tildeham(t) =
\sum_\mu\delta_\mu a^\dag_\mu a_\mu^\nodag +
\sum_{j=\textsc{a,b}} \ham[$j$](t) + \ham[\textsc{cqed}]$.

\begin{figure}
\begin{center}
\includegraphics[width=0.8\columnwidth]{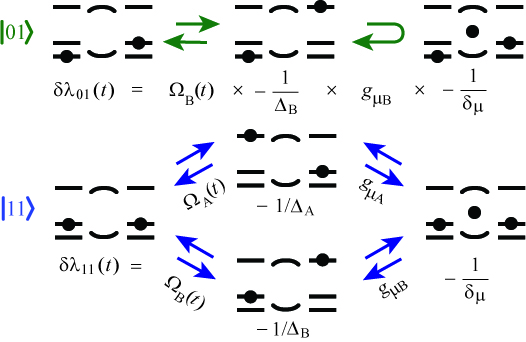}
\end{center}
\caption{(Color online) Diagrams for perturbative shifts of states $\ket{01}$ and $\ket{11}$.  The stacks of three levels indicate the quantum dots, and a cavity photon is indicated by a dot in a center.  The arrows indicate vertex terms, and the equations below indicate propagators.  A complete (fourth-order) term is a product of each vertex and propagator.}
\label{perturbation}
\end{figure}

As in the single-\QD\ case, if each $\Omega_j(t)$ evolves
slowly enough, the action of the pulse is only to cause a phase
shift on the adiabatically maintained bright states of the
system.  We will use a perturbative approach in our discussion,
although all numeric calculations use exact diagonalizations or
simulations. To fourth-order in perturbation theory, the
energies of the four qubit states shift according to the
Feynman-like diagrams of Fig. 2.  A superposition of qubit
states evolves via these energy shifts, as
\be
\ket{\psi(t)}=\sum_{j,k=\{0,1\}} c_{jk}(0)
\exp\biggl[-i\int_0^t
\lambda_{jk}(\tau)\biggr]\ket{\lambda_{jk}(t)},
\ee
where $\ket{\lambda_{jk}(t)}$ are the time-varying eigenstates
of $H$ which begin and end as separable qubit states,
$\ket{\lambda_{jk}(0)}=\ket{\lambda_{jk}(\infty)}=\tsc{\ket{j}}{a}\otimes\tsc{\ket{k}}{b}\otimes\vacket.$
An entangling gate results from the fact that the two-qubit
bright-state,
$\tsc{\ket{1}}{a}\otimes\tsc{\ket{1}}{b}\otimes\vacket$,
accrues less than twice the phase shift of either of the single
qubit bright states, such as
$\tsc{\ket{1}}{a}\otimes\tsc{\ket{0}}{b}\otimes\vacket.$  The
unitary evolution due to the pulse in this adiabatic, coherent
approximation is $U=\exp(i\tsc\theta{a}\tsc{\sigma}{a}^z)
   \exp(i\tsc\theta{b}\tsc{\sigma}{b}^z)
   \exp(i\tsc\theta{ab}\tsc{\sigma}{a}^z\tsc{\sigma}{b}^z),
$ in terms of single qubit Pauli matrices $\sigma^z_j$.  The
first two (commuting) terms are single qubit phase shifts.
These shifts may be large, as in demonstrated single-qubit
rotations
\cite{shamU1,flattespinrot,fast_rotations_prop,awschsingle,pressnature}.
The entanglement results entirely from the smaller nonlinear
phase $\tsc\theta{ab}$, which may be found in fourth-order as
\be
\label{thetaab}
\begin{split}
\tsc\theta{ab} &=
    -\int [\lambda_{11}(t)+\lambda_{00}(t)-\lambda_{01}(t)-\lambda_{10}(t)]dt
\\
    &= 2\Re{\sum_\mu\int dt\frac{\tsc\Omega{a}^\nostar(t)\tsc\Omega{b}^*(t)
                                  g_{\textsc{a}\mu}^\nostar g_{\textsc{b}\mu}^*}
        {\tsc\Delta{a}\tsc\Delta{b}\delta_\mu}} + O(g^6).
\end{split}
\ee
Note that there is no need for the \QDs\ to be resonant with
each other for this phase shift to occur, a critically important feature for inhomogeneous, self-assembled \QDs.

This gate also has a geometric interpretation
interpretation~\cite{Zheng_gate,qubus}, rendering it similar to
fast geometric phase gates enacted with trapped ions and their
phonon modes~\cite{geometric_ion}.  For this interpretation, we
note that for qubits projected into states $j$ and $k$, the
field of mode $\mu$ is nearly classical with electric field
amplitude proportional to $ \alpha^{jk}_\mu(t)=\Tr{a_\mu
                    \tsc{\bra{j}}{a}
                    \tsc{\bra{k}}{b}
                    \rho
                    \tsc{\ket{j}}{a}
                    \tsc{\ket{k}}{b}}.
$ To second order, then, two of these amplitudes are
\begin{align}
\alpha^{10}_\mu(t)&=\alpha_\mu(t)+
    \sum_\nu\frac{\tsc{g}{a$\nu$}^\nostar\tsc{g}{a$\mu$}^*}
                 {\tsc{\Delta}{a}\delta_\mu}
                 \alpha_\nu(t) + O(g^4)\\
\alpha^{11}_\mu(t)&=\alpha_\mu(t)+
    \sum_{j\nu}\frac{{g}_{j\nu}^\nostar{g}_{j\mu}^*}{{\Delta}_{j}\delta_\mu}\alpha_\nu(t)+O(g^4).
\end{align}
We may interpret this by indicating that the cavity modes are
shifted by the altered dispersive response of the cavity, depending on how many \QDs\ are loading it.  As the cavity field grows and then shrinks in time, a geometric phase accrues of size
\be
\phi^{jk}=\int_{-\infty}^{\infty}
\sum_\mu\Im{\dot\alpha^{jk*}_\mu(t)\alpha^{jk}_\mu(t)},
\ee
i.e. the \emph{area} enclosed by each phase-space path. In the
limit of large detunings $\delta$ where
$\Im{\dot\alpha^{*}_\mu(t)\alpha^{\nostar}_\mu(t)}\approx
\delta_\mu|\alpha_\mu(t)|^2$, this gives the same result, i.e.
that $\phi^{11}+\phi^{00}-\phi^{01}-\phi^{10}=\tsc\theta{ab},$
as given by \refeq{thetaab}.   This geometric picture provides
interesting perspective on the role of cavity loss.   One may
readily show in this picture that $\tsc{\bra{j}}{a}
                    \tsc{\bra{k}}{b}
                    \rho
                    \tsc{\ket{j'}}{a}
                    \tsc{\ket{k'}}{b}$
decays due to cavity loss as
$\exp[-\int^t\sum_\mu\Gamma^{jk,j'k'}_\mu(\tau)]$, where
\be
\Gamma^{jk,j'k'}_\mu(t) = \frac{\kappa_\mu}{2}
|\alpha_\mu^{jk}(t)-\alpha_\mu^{j'k'}(t)|^2,
\ee
i.e. the \emph{distance} between different paths.  This
decoherence occurs due to ability of the environment to detect
the qubit states from the light leaking from the cavity.  A
high detuning assures that state-dependent paths in phase space
are very close to each other to minimize decoherence.  If one
wants to measure the qubit states using light leaked from the
cavity, $\delta=0$ is the best strategy~\cite{hybridnjp}; if
one wants to protect the qubit states, a high value of $\delta$
allows the inherent quantum uncertainty of coherent states of
light to hide the qubit states.

Decoherence results from both cavity loss (at rate
$\kappa_\mu$) and spontaneous emission from the \QDs\ (at rate
$\gamma$).  We may roughly estimate these effects with the
following argument, in which we drop qubit subscripts for
brevity.   The dressed states of the system due to the pulse
have a trion component with probability $(\Omega/\Delta)^2$ in
first order, and they have a cavity photon with probability
$(\Omega/\Delta)^2(g/\delta)^2$ in second order.  Therefore, an
off-diagonal term of the density matrix $\tilde\rho(t)$ decays
at a combined exponent $\Gamma(t)$ of approximately
\be
\Gamma(t) \approx x\gamma \frac{|\Omega|^2(t)}{\Delta^2} +
y\sum_\mu\kappa_\mu
\frac{\Omega^2(t)|g_\mu|^2}{\Delta^2\delta_\mu^2}.
\ee
The unknowns $x$ and $y$ are in place to remind us that there
are other, order-unity constants in place depending on the
particular coherence studied. Regardless, optimal gate
operations \emph{must} work as a trade-off between the two
decoherence terms.  In the limit of large cavity-pulse detuning
$\delta$, taken as much larger than the cavity bandwidth (so
$\delta_\mu\approx \delta$), the rate of growth of the
nonlinear phase divided by the rate of decoherence,
$\Gamma^{-1}(t)\tsc{\dot\theta}{ab}$, has a maximum with
respect to $\delta$ at $\delta\propto \sqrt{\sum_\nu
\kappa_\mu|g_\nu|^2/\gamma}.$ If we tune the laser in order to
set $\delta$ at this value, and assure that $\int
\Omega^2(t)dt$ is large enough that $\tsc\theta{ab}=\pi/4$,
then off diagonal terms decay to roughly $\exp[-\int
\Gamma(t)dt]\approx \exp[-(\pi/2)\sqrt{xy/C}]$, where where
$C=4\sum_\mu |g_\mu|^2/\gamma\kappa_\mu$ is the
\emph{cooperativity} of the \QD/cavity system.  The fidelity of
the gate is then approximately
$\sim[1/2][1+\exp(-1/\sqrt{C})]$. Note that \QD\ inhomogeneity
is not critical for determining the gate fidelity.  The
critical figure of merit, the cooperativity factor $C$, goes as
the quality factor of the cavity, $Q$, divided by its mode
volume $V$.  A high fidelity gate uses a laser detuned
$\sqrt{C}$ cavity-bandwidths from the cavity resonance; for a
large-area \QD\ in a semiconductor planar microcavity, $C\sim
100$ is reasonable, in which case the optimal detuning is about
10 cavity linewidths from the central resonance.

The analysis so far has assumed the conditions allowing
adiabatic evolution.  For the present problem, in the case of
detunings $\delta$ and $\Delta$ large in comparison to the
cavity bandwidth and to Rabi frequencies $g_{j\mu}$, these
conditions are held if $\dot\Omega_j(t)\sim\sum_\mu
g_{j\mu}F_\mu(t) \ll \Delta_j^2$. Again assuming the amplitude
of $F_\mu(t)$ is chosen large enough to create a controlled-$Z$
gate ($\tsc\theta{ab}=\pi/4$) with $\delta_\mu$ at its optimum
value, and assuming a Gaussian pulse shape
$F_\mu(t)\propto(2\pi\sigma^2)^{-1/4}\exp(-t^2/4\sigma^2)$,
this condition requires $\sigma \gg \sqrt{C}
{\kappa^2}/{\Delta^2\gamma}.$ Therefore, adiabaticity may
always be obtained for long enough pulses and large enough
values of $\Delta$.  Typical pulse lengths may be in the range
of ten to hundreds of nanoseconds, with \QD\ detunings $\Delta$
on the order of THz.  This high speed is the key advantage of
self-assembled \QDs\ over similar gates enacted with trapped
ions or superconducting qubits.

\begin{figure}
\begin{center}
\includegraphics[width=0.7\columnwidth]{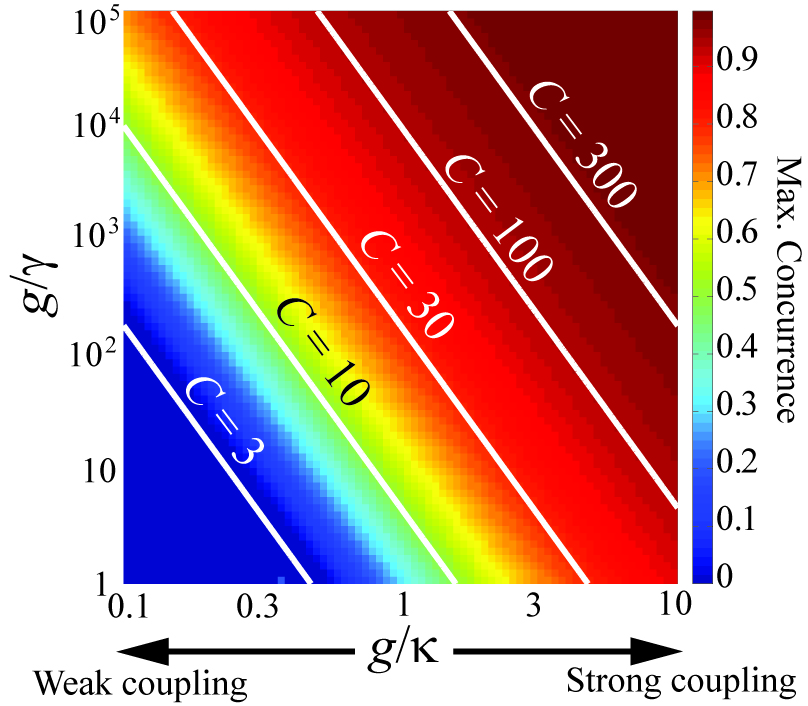}
\end{center}
\caption{(Color online) Concurrence of spins initially separable spins entangled by a gate with optimum $\delta$, calculated numerically under the adiabatic approximation $(\sigma\rightarrow\infty)$.}
\label{conarray}
\end{figure}

Evaluating the performance of the gate more rigorously than the
above approximations requires a careful figure of merit, since
as parameters are changed, the single-qubit phase shifts change
much faster than the nonlinear phase shift.  These single-qubit
phase shifts can be eliminated by single-qubit $\pi$ pulses,
and therefore should not be considered to inhibit the fidelity
of the gate; however as they are not known a-priori, evaluation
of fidelity against an ``ideal" gate with one particular choice
of phase is inappropriate.  Instead, we consider two qubits
initialized into the unentangled superposition state
$[\tsc{\ket{0}}{a}+\tsc{\ket{1}}{a}]\otimes[\tsc{\ket{0}}{b}-\tsc{\ket{1}}{b}]/2$
and evaluate the entanglement concurrence. We then test our
theory with a series of numeric approaches. In the first, we
calculate the optimal value of $\Omega_j(t)$ using
\refeq{thetaab}, assume the adiabatic theorem holds, and
evaluate the phase-shift and decay of each element of
$\tilde\rho(t)$.  This is accomplished via complete
diagonalization of $\tilde{H}(t)$ at each time $t$, and
tracking evolution of diagonal terms of standard master
equations for optical loss and spontaneous emission in the
eigenvector basis. This procedure quickly estimates gate
performance over a broad range of parameters, allowing a
numeric search for the value of $\delta$ which maximizes the
concurrence.  The maximum concurrence resulting from this
parameter search is shown in \reffig{conarray}, which verifies
that the figure of merit is indeed $C$.  To test further, we
perform detailed solutions of a complete master equation.  No
adiabatic or perturbative approximations are made; the time
evolution of $\tilde{\rho}(t)$ is explicitly integrated using
Runge-Kutta or semi-implicit extrapolation integration
techniques.  Again, only a single cavity mode is assumed, but
it may admit more than one photon.   These simulations are
slower, but verify the adiabatic theory for long enough pulse
lengths.

Realistically, it is unlikely that the adiabaticity of the
system will limit the pulse length or the allowable detunings
and homogeneity ($\tsc\Delta{a}$ and $\tsc\Delta{b}$).  As
these detunings are increased, the amount of optical power
required to complete a gate increases as well, and eventually
additional decoherence sources not modeled here will arise.  In
InAs \QDs\ and donor impurities in GaAs, such additional
decoherence is seen in studies of single qubit rotations, in
many cases preventing high fidelity pulses of large angles
\cite{awschsingle}, but in other cases allowing them
\cite{pressnature}.  The optimization of sample parameters to
balance \QD\ inhomogeneity and allowable optical power will be
the critical hurdle for the successful demonstration of this
gate in a semiconductor system.

We now address the type of small-mode-volume cavity to be used
for this gate.  Transmission line resonators in circuit circuit
cavity \abbrev{qed}~\cite{wiringup}, photonic crystal
cavities~\cite{fushman08}, and whispering gallery modes of
microdisks~\cite{gerard05a} are possibilities, and if such
cavities are critically coupled to waveguides and to other
cavities to form complex ``photonic molecules," large-scale
quantum computer architectures relying on this gate may be
constructed~\cite{fast_rotations_prop,ijqi}.  However, the
large-scale integration of these cavity \abbrev{qed}\ systems
present a variety of challenges. A promising, simpler route for
scalability, enabled by a large-area \QD\ lattice, is
integration with a planar microcavity sample, as shown in
\reffig{levels}(c). The two-qubit gate would function by
exciting a laser spot with the inherent cavity mode size
$\tsc{r}{C}=\lambda_0/\sqrt{2\pi(1-R)}$, where $\lambda_0$ is
the optical wavelength in vacuum and $R=\sqrt{R_1R_2}$ is the
effective reflectivity~\cite{bhy93}, which overlaps two
neighboring, site-controlled \QDs\ in a two dimensional
array~\cite{forchel_alignment}. The cooperativity factor of a
\QD\ in a planar microcavity is not the same as that of a point
dipole; instead it depends on the size of the \QD.  The
mode-volume $V$ may be estimated under the optimal condition
that the angular distribution of the density of photon states
of the cavity match the dipole emission lobe of the
\QD~\cite{bpjy94}.  In this case $4\sum_\mu
|g_\mu|^2/\kappa_\mu/\gamma  \approx Q\lambda_0^3/ [\pi^3
(\ts{a}{b}^*)^2 L],$ where $L$ is the cavity length. This
approximation is valid for \QDs\ large in comparison to the
optical wavelength in the semiconductor.  High values of
concurrence are expected to be possible with existing quality
factors of planar microcavities \cite{reitzenstein07} and
site-controlled, large-area quantum
\QDs~\cite{forchel_alignment}.

As a final note, the present analysis may be readily
generalized to more complex level structures.  For example, if
a full $\Lambda$-system is employed with a small splitting
between the ground states, a single pulse may still enact a
gate, but now it will include qubit flip-flop terms, similar to
two-laser cavity \abbrev{QED} gates~\cite{IABDLSS99}.  The key
new advantage of the present proposal is its scalability to as
many as $10^9$ qubits in a single wafer and massively parallel
creation of cluster states by simultaneous, fast two-qubit
gates, with high tolerance for large optical \QD\
inhomogeneities.  The relative simplicity and parallelism of
this two-qubit gate may allow several possible architectures
for fast and large-scale optically controlled quantum
computers.

We acknowledge fruitful discussions with Bill Munro, Kae
Nemoto, and Sebastien Louis.  This work was supported by the
National Science Foundation CCR-08 29694, NICT, and MEXT.



\end{document}